\documentclass[conference]{IEEEtran}
\usepackage[OT1]{fontenc}
\IEEEoverridecommandlockouts
\usepackage{cite}
\usepackage{amsmath,amssymb,amsfonts}
\usepackage[linesnumbered,ruled,vlined]{algorithm2e}
\usepackage{graphicx}
\usepackage{textcomp}
\usepackage{xcolor}
\usepackage{amsmath}
\usepackage{booktabs} 
\usepackage{stfloats}
\usepackage{array}
\usepackage{multirow}
\usepackage{subfigure}
\usepackage{graphicx}
\usepackage{subfig}
\usepackage{caption}
\usepackage{enumitem}
\setlist[itemize]{leftmargin=2em}
\setenumerate[1]{itemsep=0pt,partopsep=0pt,parsep=\parskip,topsep=5pt}
\setitemize[1]{itemsep=0pt,partopsep=0pt,parsep=\parskip,topsep=5pt}
\setdescription{itemsep=0pt,partopsep=0pt,parsep=\parskip,topsep=5pt}
\graphicspath{{Figures/}{pictures/}{images/}{./}} %

\def\BibTeX{{\rm B\kern-.05em{\sc i\kern-.025em b}\kern-.08em
		T\kern-.1667em\lower.7ex\hbox{E}\kern-.125emX}}
\begin{document}	
	\label{tiltle}
	
	\title{Towards Holographic Video Communications: A Promising AI-driven Solution}
	
	\label{author}
	\author{
		
		Yakun~Huang,
		Yuanwei~Zhu,
		Xiuquan~Qiao,
		Xiang~Su,~\IEEEmembership{Member,~IEEE,}\\
		Schahram~Dustdar,~\IEEEmembership{Fellow,~IEEE,}
		Ping~Zhang,~\IEEEmembership{Fellow,~IEEE}
		\IEEEcompsocitemizethanks{
			\IEEEcompsocthanksitem  Y. Huang, Y. Zhu, X. Qiao and P. Zhang are with the State Key Laboratory of Networking and Switching Technology, Beijing University of Posts and Telecommunications, Beijing 100876, China. E-mail:\{ykhuang, zhuyw, qiaoxq, pzhang\}@bupt.edu.cn. (X. Qiao is the corresponding author.)
			\IEEEcompsocthanksitem X. Su is with the Department of Computer Science, Norwegian University of Science and Technology, 2815 Gj{\o}vik, Norway and University of Oulu, 90570, Oulu, Finland. Email:xiang.su@ntnu.no.
			\IEEEcompsocthanksitem S. Dustdar is with the Distributed Systems Group, Technische Universität Wien, 1040 Vienna, Austria. E-mail:dustdar@dsg.tuwien.ac.at.
		}
	}
\maketitle
\label{abstract}
\begin{abstract}
	Real-time holographic video communications enable immersive experiences for next-generation video services in the future metaverse era. However, high-fidelity holographic videos require high bandwidth and significant computation resources, which exceed the transferring and computing capacity of 5G networks. This article reviews state-of-the-art holographic point cloud video (PCV) transmission techniques and highlights the critical challenges of delivering such immersive services. We further implement a preliminary prototype of an AI-driven holographic video communication system and present critical experimental results to evaluate its performance. Finally, we identify future research directions and discuss potential solutions for providing real-time and high-quality holographic experiences.
\end{abstract}

\section{Introduction}
The holographic video provides users with an immersive six degrees of freedom~(6-DoF) viewing experience than traditional virtual reality~(VR), 360-degree, and other 3-DoF videos~\cite{liu2021point}.
6-DoF videos are characterized by having depth information for each frame, providing 3-DoF of translational movement (X, Y, Z) and 3-DoF of rotational movement (yaw, pitch, roll).
6-DoF videos allow users to walk around an object in a circle and view it from the top and the bottom.
Point cloud video~(PCV), as a representative holographic 6-DoF video service, describes the objects using a set of disordered 3D points with coordinates and color.
Figure~1 compares the PCV transmission with different video services.
PCV stream (e.g., capturing one second of raw PCV with one depth camera at 30 frames per second (FPS) produces 2.06~Gb of data) is highly time- and resource-consuming for encoding and decoding, requiring at least an hour for a common computer compared with 3-DoF videos.

More importantly, PCV transmission requires a bandwidth capacity of more than Gbps level, far beyond the current transmission capacity of 5G networks.
Undoubtedly, the holographic video introduces requirements far exceeding traditional video streaming services regarding network bandwidth, transmission latency, and computing complexity.
\begin{figure*}[htbp]
	\centering
	\includegraphics[width=\textwidth]{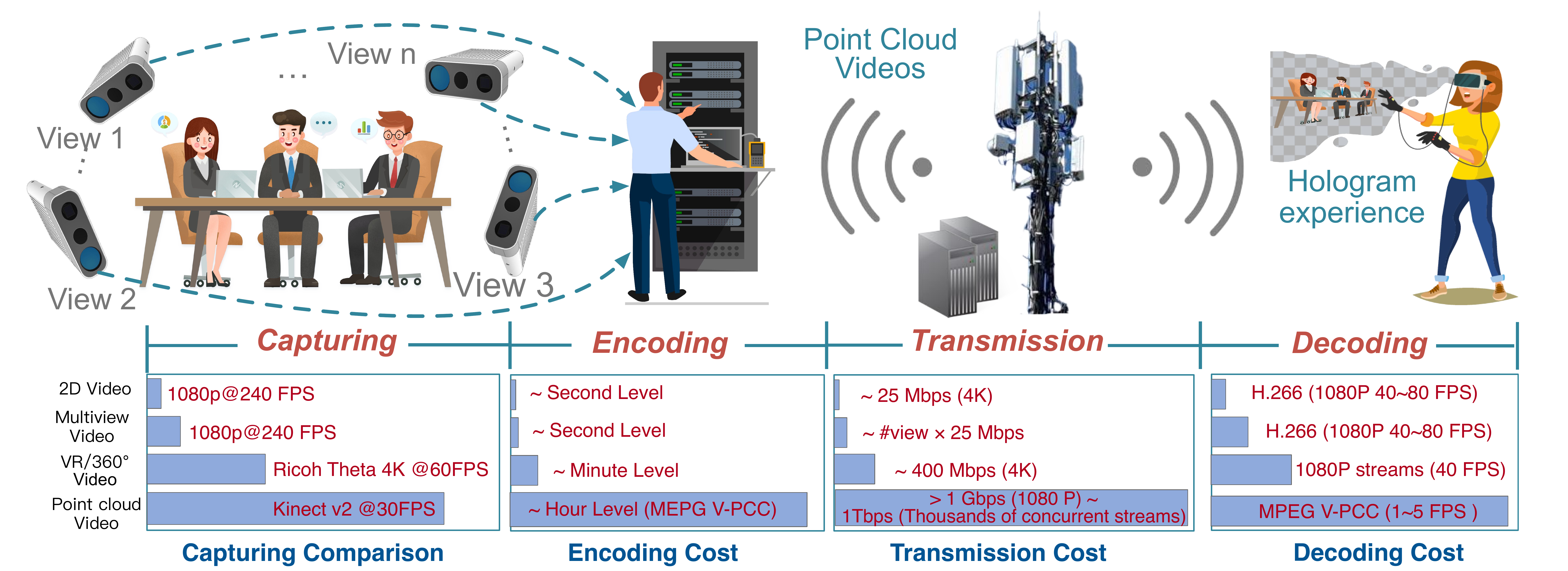}
	\caption{Comparisons between holographic point cloud video with conventional videos.}
	\label{Fig_1}
\end{figure*}

We investigate the transmission techniques for PCV, including point cloud compression and video streaming optimization.
For compression, traditional methods include Kdtree-based and Octree-based solutions, such as the popular Point Cloud Library~(PCL)~\cite{Rusu_ICRA2011_PCL} and Draco~\cite{draco}.
ISO/IEC Moving Picture Experts Group (MPEG) standardizes Video-based Point Cloud Compression(V-PCC) and Geometry based Point Cloud Compression~(G-PCC) for PCV.
However, these methods require higher computing resources and costs than 3-DoF videos.
Besides, although some deep learning-based compression techniques provide lower accuracy loss and higher compression ratios~\cite{qi2017pointnet++,li2019pu}, they are only applicable for offline holographic video pre-processing due to high computing overhead and inference latency.
For video streaming optimization, most point video steaming techniques expand 3-DoF video streaming methods such as tiling and view angle prediction.
Since PCV adds extra 3-DoF information, it requires more adaptive adjustment of streaming than 3-DoF with the dynamic change in the physical distance between the user and the scene.
Some research investigates the combination of point cloud compression and transmission optimization~\cite{han2020vivo, wang2021qoe}.
For streaming quality of service~(QoS) management, Zhang \textit{et al.}~\cite{zhang2019covering} proposed a covering-based quality prediction method to accurately predict the QoS, along with the query of quality correlation~($Q^2C$) model~\cite{zhang2018efficient} for the QoS guarantee.
However, these solutions cannot be run in real-time on mobile devices due to the massive cost of video compression and codecs.

We review related surveys, tutorials, and magazine publications on PCV, holographic video, and immersive video.
Liu~\textit{et al.}~\cite{liu2021point} discuss the challenges and solutions to adaptive point cloud streaming and provide a prototype of extending MPEG Dynamic Adaptive Streaming over HTTP~(DASH).
Clemm~\textit{et al.}~\cite{clemm2020toward} articulate the networking challenges to enable immersive holographic videos and propose new network architecture for optimizing the coordination and synchronization of concurrent streams.
Hooft~\textit{et al.}~\cite{van2020capturing} present the status and challenges of 6-DoF media and Taleb~\textit{et al.}~\cite{taleb2021extremely} provide an overview of immersive services as well as the relevant industry and standardization activities.
These works highlight the gap between existing streaming solutions and implementing PCV transmission.
Most solutions extend from 3-DoF video compression or adaptive streaming techniques and fail to involve an AI-native PCV streaming.

This article introduces the landscape and requirements of holographic PCV communication and analyses the technical challenges associated with supporting PCV services.
We propose an advanced AI-driven transmission solution as a prototype for preliminary exploration and verify its performance.
Our contributions pertain to 1) a novel transmission mechanism for holographic PCV; 2) the end-to-end network design that joins the encoder and decoder; and 3) adaptive streaming technology for proposed AI-driven video transmission.
We further discuss the proposed AI-driven communication technique and identify future directions for high-quality PCV services.

\section{Requirements and Challenges}
\subsection{Requirements}
Holographic PCV streaming poses significant demands on network transmission infrastructure in terms of ultra-low delay and reliable network, heavy computation, and device mobility and portability.

\subsubsection{Ultra-low delay and reliable networks.} 
6-DoF movement and orientation features of PCV are more sensitive to delay than 3-DoF video services, whose ideal delay requirement is less than 5~ms and is more stringent than that of traditional 3-DoF videos~(i.e., $<$20~ms)~\cite{strinati20196g}.
Since PCV requires numerous depth cameras to capture data, this further increases data volume than other types of videos.
Therefore, continuous and reliable transmission of multi-view captured PCV streams requires a reliable network and lower network jitter than 3-DoF video transmission.

\subsubsection{Heavy computation.}
Encoding and decoding a PCV using the MPEG standard are computationally intensive, even if we ignore the computations used for capturing.
For example, encoding a one-second video from the \textit{longdress} dataset with lossy compression requires 11 to 42 minutes using MPEG V-PCC on a generic computer \cite{lee2020groot}.
Although we can use high-performance GPU servers to accelerate the encoding of PCV at the sender, the computing capability of mobile devices, such as AR/VR glasses, does not fulfill the requirements of real-time decoding.
Thus, massive encoding and decoding computation requirements are one of the primary reasons that hinder the provision of a 6-DoF experience on mobile devices.

\subsubsection{Device mobility and portability.}
Holographic PCV introduces higher demands on device mobility and portability.
We can use cable-connected VR terminals or large display screens to enable immersive experiences in panoramic and 360-degree VR videos, which is currently one of the primary methods of immersive interactions.
However, PCV will significantly reduce the 6-DoF experience if users are not free and flexible to move and interact.
Therefore, interactive devices for holographic PCV need to provide free mobility.
Portable devices are crucial to providing a satisfying holographic PCV experience.

\subsection{Challenges}
\subsubsection{\textbf{Disordered point cloud points and massive computing demands challenge the traditional streaming pipeline.}}
Point clouds are represented by massive disordered 3D points (X, Y, Z) and colors (R, G, B). 
It requires hundreds of thousands of points to clearly represent 6-DoF contents, which makes the data volume of PCV much larger than that of 3-DoF videos.
In addition to the intuitive increase in data volume, we have to address the challenges of compression, encoding, and decoding for real-time PCV transmission.
However, existing encoding and decoding methods extending MPEG standards are mainly for offline video services, which cannot provide real-time decoding on mobile devices.
Although some AI-based compression techniques extract point cloud features and acquire a better compression rate than traditional methods, they require extensive GPU resources.
Also, they have to train another heavy neural network to reconstruct the original point cloud, which also cannot provide real-time decoding for PCV transmission.
Hence, there is no existing end-to-end lightweight AI network designed for point cloud transmission from the original point cloud to the final rendering point cloud.

\subsubsection{\textbf{Intensive 6-DoF point cloud decoding challenges resource-constrained mobile devices.}} 
Compared with the existing mature 3-DoF VR or 360-degree video, 6-DoF PCV lacks efficient decoding algorithms. 
The computing capability and mobile energy consumption limit the usage of existing decoding solutions, especially for resource-constrained devices.
This introduces that the existing 6-DoF PCV has to compromise using 3-DoF delivery and only providing a VR video-like service.
When using AI-based methods to extract key features for transmission, running real-time model inference for reconstructing point clouds on mobile devices is a significant challenge.
Currently, the AI-based reconstruction method for point clouds such as PU-GAN~\cite{li2019pu} requires intensive computation and GPU resources to support a satisfying experience on the high-performance server.
Therefore, implementing AI-based decoding on mobile devices is still a big challenge for 6-DoF PCV.

\subsubsection{\textbf{Adaptive delivery challenges AI-driven video streaming.}} 
6-DoF PCV can initially provide adaptive delivery by extending MPEG DASH and dynamic tiling strategies of 3-DoF videos.
Also, some efforts apply deep reinforcement learning~(DRL) algorithms to improve transmission performance.
However, these methods provide adaptive streaming based on a traditional video transmission pipeline, including compression, encoding, transmission, and decoding.
Such adaptive methods are difficult to be applied to AI-driven methods that extract key features for transmission.
More importantly, the traditional adaptive streaming mainly considers the network environment, while the decoding involves intensive computation on the mobile device.
Thus, incorporating the device computing capability into adaptive streaming delivery and designing an appropriate DRL is a significant challenge for AI-driven PCV transmission.

\section{Progressive real-time delivery: A novel AI-driven Solution}
\begin{figure*}[htbp]
	\centering
	\includegraphics[width=\textwidth]{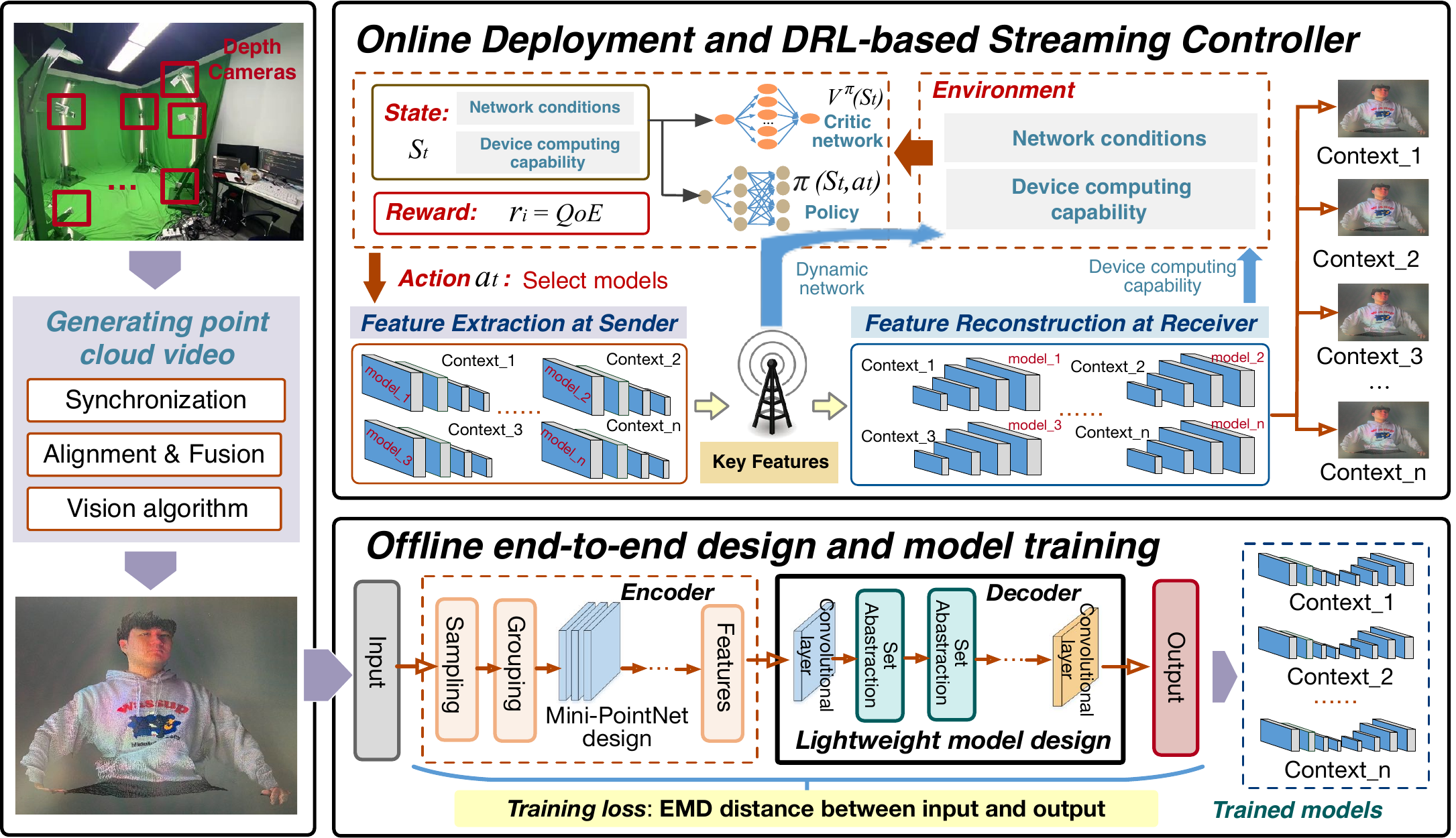}
	\caption{Proposed AI-driven end-to-end transmission architecture.}
	\label{Fig_2}
\end{figure*}
We propose a novel AI-driven solution that makes PCV streaming as an end-to-end neural network training problem.
Existing end-to-end designs for point clouds either extract key feature points from the original point cloud for object detection or complete the point cloud with the generative adversarial network (GAN) technique.
Unlike those solutions, we radically design the point cloud feature extraction and reconstruction as an end-to-end process for training to obtain better feature extraction capability and reconstruction results.
\textit{To address the first challenge}, we design the encoding and decoding process as an end-to-end trainable neural network and transfer the encoder's key features instead of the compressed point cloud.
Based on this design, AI-driven PCV delivery evolves from the traditional pipeline to use trained point cloud feature extraction for encoding and point cloud reconstruction for decoding.
\textit{To address the second challenge}, we first optimize the point cloud downsampling process, which has a high computational complexity in the reconstruction of decoding. 
Then, we design a pruning and quantization joint method to reduce model parameters and size and speed up online decoding for mobile devices.
\textit{To address the third challenge}, we propose an adaptive control method for dynamic AI-driven point cloud transmission. 
By sensing dynamic contexts, we propose a DRL-based adaptive transmission method with user QoE as the optimization goal, considering the transmission latency and reconstruction accuracy.
It can adaptively and dynamically match the optimal point cloud encoder-decoder models to obtain the optimal transmission experience according to the used mobile devices and the network condition.

Figure~2 presents the overall AI-driven transmission architecture for PCV service, consisting of PCV generation, training end-to-end encoder-decoder models, online transmission, and adaptive streaming.
\subsubsection{\textbf{Point cloud video generation.}} 
We first implement a point cloud generation to get a real-time PCV stream.
As shown in Figure~2, we deploy multiple depth cameras at different angles to capture PCV frames.
Then, we convert them to point clouds and fuse them into a complete 6-DoF PCV stream with generative learning algorithms, which is out of the scope of this article. Instead, we focus on how to provide efficient and adaptive transmission services for PCV.

\subsubsection{\textbf{Offline end-to-end design and model training.}}
As shown in Figure~2, the proposed end-to-end encoder-decoder design takes the original point cloud as input and outputs a reconstructed point cloud identical to the original point cloud. This AI-driven mechanism transmits key features instead of compressed PCV, reducing redundant data transmission. Then, it uses a lightweight and efficient model to reconstruct the point cloud on mobile devices. For the design of the encoder, we use the hierarchical extraction structure of PointNet++~\cite{qi2017pointnet++} and employ an ensemble abstraction layer to capture the local structure from the original point cloud. In detail, we use several \textit{Sampling} layers, \textit{Grouping} layer, and a \textit{Mini-PointNet} layer to encode local region patterns into feature vectors. Then, the raw input frame is represented by fewer points and features when outputting a point-by-point feature matrix. For the decoder, since the conventional GAN-based reconstruction model has a large model size and a long inference time on mobile devices, we cannot directly use the GAN-based approach as the design of the decoder. To reduce the model size and improve the inference efficiency of the decoder, we propose a lightweight GAN-based point cloud reconstruction network using the model pruning and quantification techniques. 
We start with using the generator of the PU-GAN network as the backbone of the decoder. We generate more diverse point distributions to enhance the feature variations rather than a simple duplication strategy by introducing the upsampling-downsampling-upsampling layer~\cite{li2019pu}.
Then, we join the encoder with the above basic decoder for end-to-end training and obtain a basic encoder-decoder model with optimal accuracy. Then, we perform a weight pruning operation on the trained decoder model and then propose an 8/16-bit quantization acceleration. Finally, we jointly fine-tune the parameters with the encoder during pruning and quantization to obtain the optimal lightweight decoder model. Note that we do not aggressively use the binary quantization technique because we need to guarantee the reconstruction accuracy as much as possible.

To train the designed novel neural network, the repulsion loss and the uniform loss commonly used are not effective~\cite{li2019pu}.
This is because the repulsion loss avoids the generated points near the original points and the uniform loss ensures generating point sets in a uniform distribution.
However, our AI-driven transmission network aims to reduce the distance between the original and generated point cloud as much as possible.
Thus, we use the earth mover's distance~(EMD) as the loss function to produce generated points on the target surface, which are similar to the original input. 
EMD can measure the distance between the original and generated point cloud distributions.
We decompose the original point cloud into 200 patches and employ the patch-based training strategy for input with a large number of points.

It is important to provide lightweight models under various contexts, such as network conditions and different mobile devices.
This also means that such an AI-driven approach requires adaptive online delivery streaming techniques.
Hence, we train and cache encoder-decoder models for different contexts offline to match different device computing capabilities and network conditions. 
Also, we can retrain and update encoder-decoder models as the context changes.

\subsubsection{\textbf{Online transmission and streaming controller.}}
The online adaptive streaming controller requires choosing the optimal encoder-decoder model for dynamic contexts, including the network condition and the device computing capability. For each trained encoder-decoder model, we define a hyperparameter to represent the compression ratio between the original point cloud frame and the feature vector output by the encoder. Hence, the online adaptive streaming controller aims to select the best encoder-decoder model by adjusting different hyperparameters to obtain the maximum compression ratio while satisfying the accuracy of the reconstructed PCV. To implement adaptive streaming for the proposed AI-driven transmission, we propose an online self-learning controller based on DRL, providing optimal encoder-decoder model selection for dynamic contexts. As shown in Figure~2, we construct the self-learning streaming controller by taking the current network condition, the device computing capability, and the demands as the state, defining the reward, and selecting the encoder-decoder model as the action for DRL policy network training. We show the state, the action, and the reward defined in the DRL-based online streaming controller. The reward is essential in achieving fast convergence and obtaining the optimal global solution. We use the quality of experience~(QoE) as the reward, which considers both transmission latency and reconstruction accuracy to optimize the DRL training. Typically, rate-distortion optimization~(RDO) is used to measure and optimize the compression performance for traditional videos or images. RDO considers the performance of the lossy (image quality) and bitrate (the amount of data required to encode). Similarly, this is also useful for compressed transmission of PCV. Hence, we define the QoE from the reconstruction accuracy and transmission time, representing distortion rate and bit rate, respectively. This also means the smaller the amount of data, the smaller the transmission time, and the smaller the FPS and delay of user experience. 
Specifically, we formulate the QoE through the F1-Score, the harmonic mean of the precision and recall, to achieve an optimal value on these two indicators.
In summary, online streaming controls the hyper-parameter according to the dynamic network and device computing capability and obtains different encoder-decoder transmission models, dynamically adjusting the compression rate on the mobile device. 
In our implementation, we consider four network conditions, i.e., 3G, 4G, WiFi, and 5G, and four levels of device computing capabilities by controlling the number of CPU cores.

\section{Experimental Analysis}
The deep neural network described in Section III is a high-level framework.
Our specific architecture adopts the hierarchical extraction module based on the backbone of PointNet++ \cite{qi2017pointnet++} as the encoder, and adopts the feature expansion component and point set generation component in the generator of PU-GAN \cite{li2019pu}. 
We first implement a basic AI-driven transmission model using the PointNet++~\cite{qi2017pointnet++} as the encoder's backbone and join the generator as the decoder.
Then, we describe the implementation of the lightweight decoder model, including the high-precision pruning and quantization.
Specifically, we adopt the weight pruning strategy to remove part of the weights from convolutional neural network layers and make the model faster and smaller.
Last, we use 8/16-bit quantization operation and fine-tune model parameters to obtain the optimal lightweight decoder model for different devices.

\begin{figure}[htbp]
	\centering
	\includegraphics[width=0.45\textwidth]{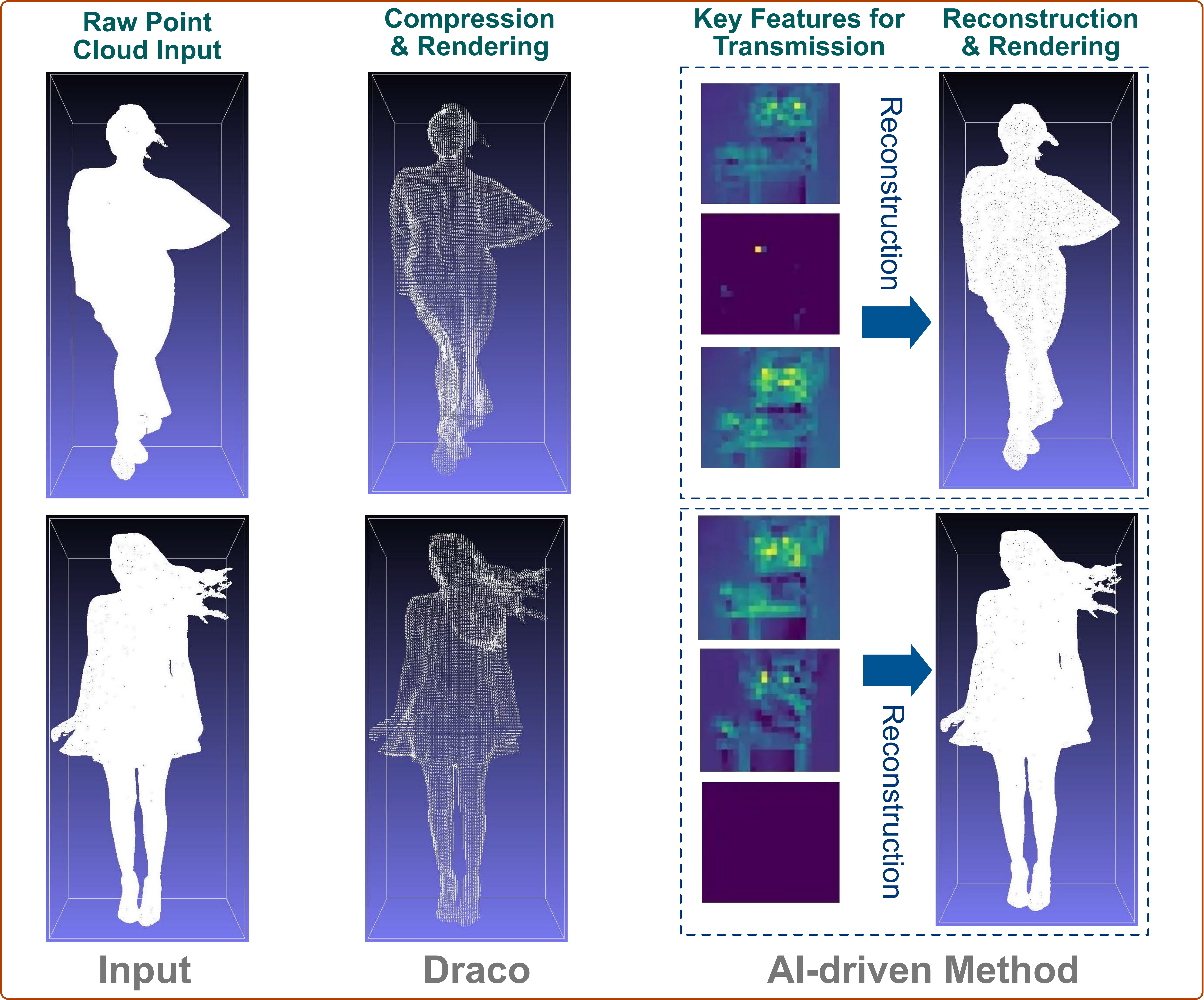}
	\caption{Qualitative comparisons on the reconstruction results.}
	\label{Fig_3}
	\vspace{-0.2cm}
\end{figure}
To improve the generalization of the AI-driven model, we train our end-to-end neural network by utilizing 147 3D point cloud objects \cite{li2019pu}.
The dataset includes a rich variety of objects, from simple objects (e.g., Icosahedron) to high-detailed objects (e.g., Statue). 
In addition, we use four real-world PCV sequences for testing~\cite{d20178i}, each of which is a human body captured by 42 RGB cameras at 30 FPS over a 10s period. 
Due to space constraints, we select the two typical longdress and redandblack datasets to show the reconstruction performance in Figure 3. 
Note that we decompose each point cloud frame into multiple patches of the same size in advance to unify the input dimensions.
In particular, we group each patch with 256 points and normalize them in a unit sphere.
Then, we compress the patch (256,3) (i.e., 256 points with 3D coordinates) into a (5,5) feature vector matrix.
We compare our method with Draco \cite{draco} by setting the compression level parameter (cl) as ten and the quantization parameter (qp) as eight. 
Figure~3 shows that Draco performs a nonuniform and ``blocky" phenomenon when compressing the raw point cloud frame by 11x and 13x.
The quantization bits are not precise enough to represent the coordinate information. 
Moreover, our AI-driven solution achieves a maximum compression ratio of 30x while ensuring impressive reconstruction results.

\vspace{-0.1cm}
\begin{figure}[htbp]
	\centering
	\includegraphics[width=0.49\textwidth]{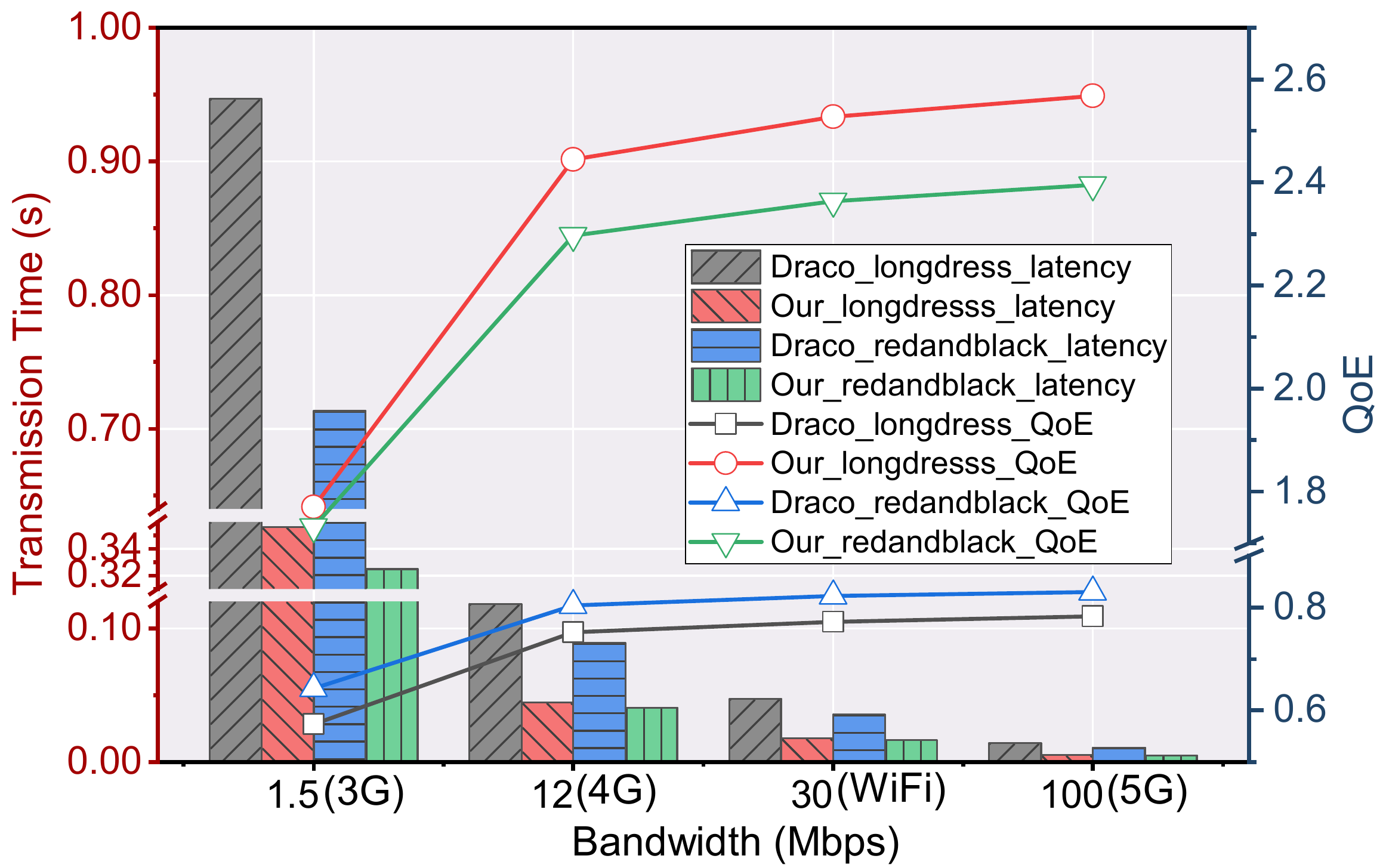}
	\caption{Quantitative comparisons of the AI-driven method with Draco. The bars denote the transmission latency, and the lines denote the QoE performance.}
	\label{Fig_4}
	\vspace{-0.2cm}
\end{figure}
We evaluate the performance of the AI-driven method with popular Draco compression.
Figure 4 presents quantitative experimental results for evaluating the transmission latency and QoE in various network conditions.
We find that transferring a point cloud frame using the proposed AI-driven framework significantly reduces latency compared with Draco. 
Meanwhile, our method achieves higher QoE than Draco on both selected datasets. 
The results illustrate the superiority and robustness of the AI-driven framework.
   \begin{figure}[htbp]
	\centering
	\includegraphics[width=0.45\textwidth]{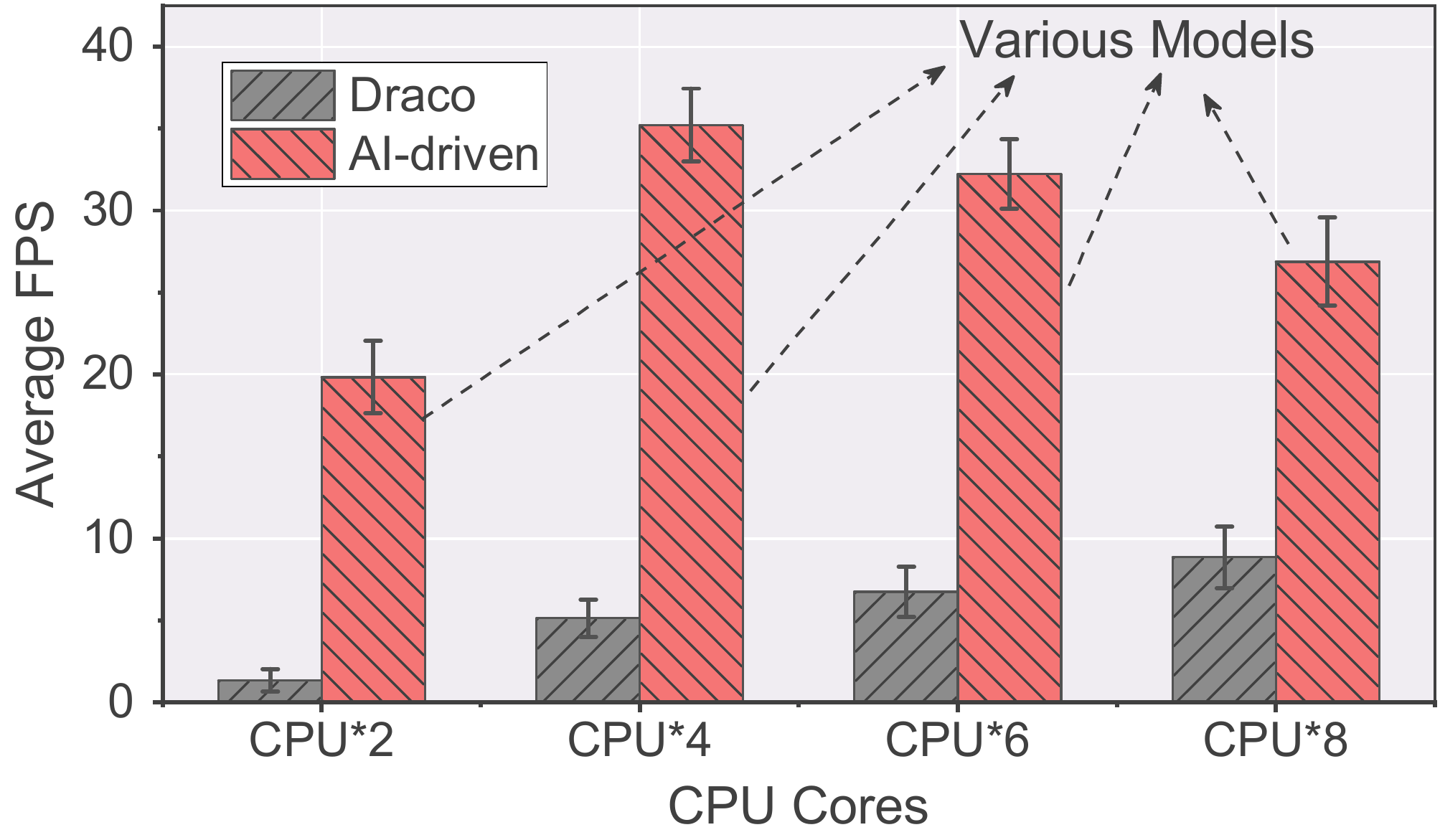}
	\caption{FPS performance of various computing capabilities.}
	\label{Fig_5}
		\vspace{-0.4cm}
\end{figure}
Besides, to verify whether the online controller can provide adaptive transmission under a dynamic network environment, we have trained several encoder-decoder models, whose transmitted feature vector matrix sizes are represented as (06, 06) to (20, 20).
We further evaluate AI-driven reconstruction with Draco under various computing capabilities in Figure~5.
Note that we set the network as WiFi, and both methods have a similar QoE performance.
The results show that our method achieves a high FPS performance without quality loss and supports real-time immersive reconstruction and rendering. 
Our approach provides different encoder-decoder models according to computing capability. In particular, AI-driven shows a significant advantage with 2 CPU cores.

\vspace{-0.2cm}  
\begin{figure}[htbp]
	\centering
	\includegraphics[width=0.48\textwidth]{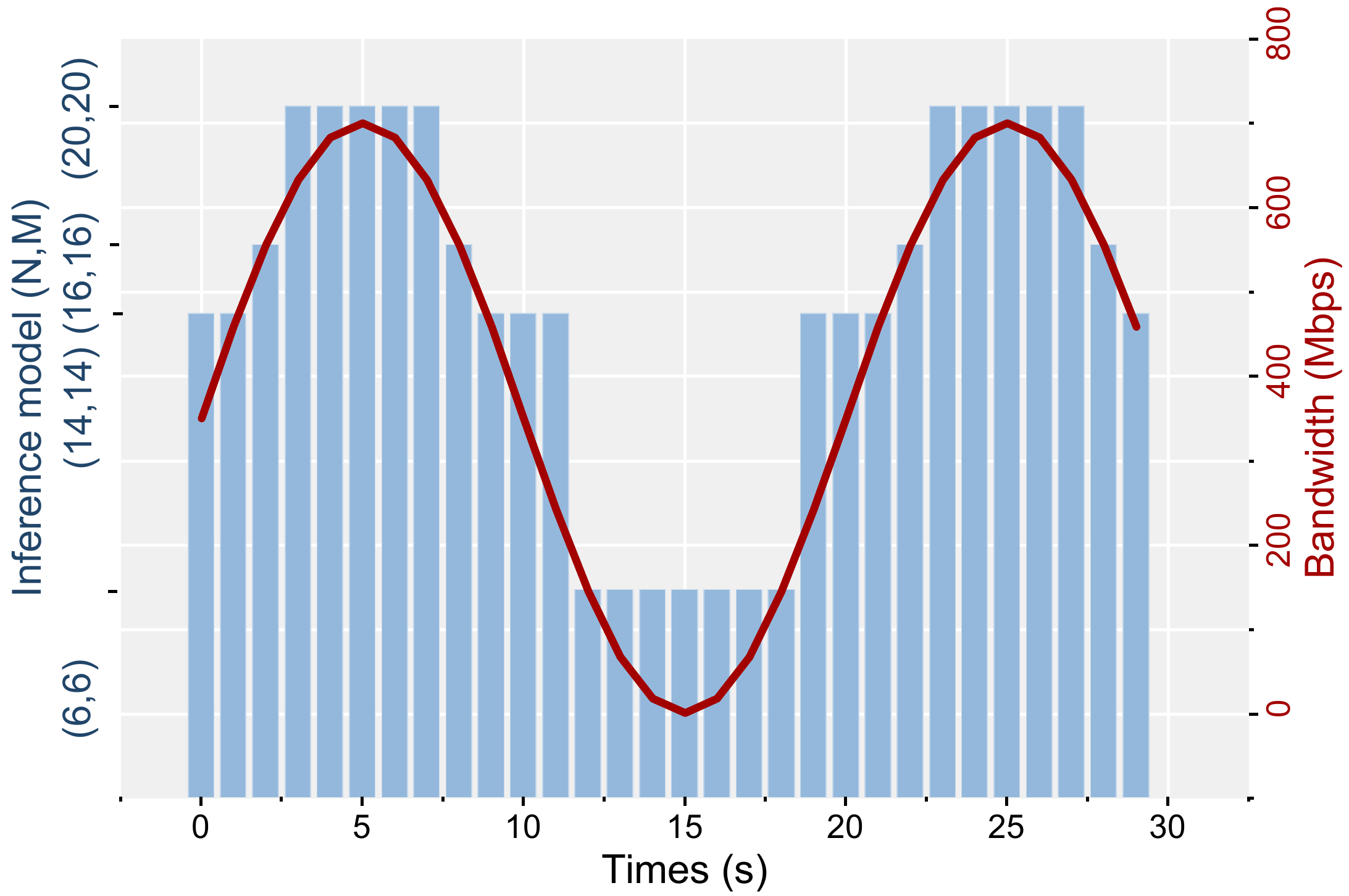}
	\caption{Performance of DRL-based online controller.}
	\label{Fig_6}
	\vspace{-0.2cm}  
\end{figure}
To evaluate the performance of the online controller, we have trained the asynchronous advantage actor-critic~(A3C) network and used the trained actor-network to select the transmission model. 
The accumulative discount rewards reach convergence at about 700 episodes in the training phase.
We also show the model selection results in the testing phase in Figure~6, illustrating the effectiveness of a DRL-based online controller.
In addition, the red curve represents the bandwidth change over time and the blue bars represent the inference models over time outputted by the online adapter. 
The adaptive adjustment of the inference models has the same trend as the dynamic changes of network conditions, this then demonstrates the effectiveness of the online adapter.

\section{Future Directions}
AI-driven point cloud streaming address fundamental challenges of efficient holographic transmission.    
We further discuss important research directions for AI-driven PCV delivery and point out potential solutions.
\subsubsection{\textbf{Interest-aware PCV capturing.}}
Real-time and high-quality PCV capturing are important in achieving holographic communication and interaction.
We provide a preliminary real-time PCV capturing system to implement the proposed AI-driven transmission.
However, existing capturing schemes cannot obtain the same video quality as a 3-DoF video with a low FPS of around 10$\sim$15. 
Also, the larger range of point cloud capturing increases pressures on network bandwidth and computation.
Encoding all captured PCV is very computationally intensive, but not all the PCV content is within the user's viewpoint and interest.
Therefore, an important future research direction is to achieve real-time, high-quality PCV capturing, especially for AI-driven transmission methods.
A potential solution can dynamically capture the area of the user's interest and the user's viewpoint instead of full-field capturing.
This can greatly improve the capturing efficiency and reduce the transmission data volume of the PCV.

\subsubsection{\textbf{Extending AI-Driven transmission with MPEG.}}
The proposed AI-driven model only considers encoding and decoding each raw input frame.
This means that two adjacent frames are transmitted similarly without considering the motion and spatial relationships between frames.
Although the AI-driven transmission method is hard to extend MPEG directly, it is a future research direction to extend the AI-driven transmission by incorporating some advantages from MPEG standards.
One potential approach is introducing keyframe and dynamic frame concepts for improving AI-driven transmission. 
For example, a high-precision encoder-decoder transmission model can be used for keyframes and dynamic frames, while a low-precision and encoder-decoder transmission model can be used for non-key and static frames.

\subsubsection{\textbf{Balance between communication, computation, and storage.}} AI-driven PCV transmission requires significant computational and storage resources by offline training of encoder-decoder models matching different contexts. Traditional adaptive bitrate streaming~(ABR) algorithms require extensive storage resources to cache videos of different resolutions for dynamic transmission. However, AI-driven transmission is more complex in adaptive streaming, consuming massive computing resources for training and executing encoder-decoder models. Also, the proposed method requires storage resources for caching offline trained models. Therefore, one future research direction is balancing communication, computation, and storage resources to optimize multidimensional network resources. Moreover, the potential approach is to make the flexible adjustment between computation and caching according to the scenarios and QoS requirements. For example, we may train fewer low-precision models when computing resources are abundant. When storage resources are abundant, we can appropriately increase training low-precision models to match weak computing capability or network conditions.

\subsubsection{\textbf{Quality assessment of AI-driven transmission.}} The AI-driven transmission method uses the QoE metrics that consider the transmission time and reconstruction accuracy. We have validated that the proposed QoE can help provide adaptive streaming and measure the performance of the AI-driven method on several datasets. 
However, it is an important direction to explore comprehensive assessment methods from involved environmental factors rather than only from the reconstruction accuracy and transmission latency.  
For example, we will study a comprehensive, objective assessment model by considering user behavior.
Also, considering the correlation between complex environmental factors such as networks, media, and the device enables a more objective quality evaluation.

\section{Conclusion}
The article reviewed the landscape of hologram video in the form of point clouds, clarified the differences between PCV with conventional videos, and revealed that existing technologies are still far from supporting real-time holographic video streaming.
We discuss the critical challenges of enabling holographic communication and providing immersive services in transmission technology, computing, mobility, and ubiquity.
We further propose a novel point cloud streaming method that is completely different from existing delivery mechanisms from an AI perspective, extracting key semantic features for delivery and rendering.
Finally, we point out some future directions to facilitate research in PCV and immersive services.

\section*{Acknowledgment}
The authors would like to thank Jacky Cao for his diligent proof-reading of this article.
This research was supported in part by the International Cooperation and Exchange of NSFC under Grant 61720106007, in part by the National Natural Science Foundation of China under Grant 62202065, in part by the Project funded by China Postdoctoral Science Foundation 2022TQ0047 and 2022M710465, and in part by Academy of Finland under projects 319670 and 326305.
	
\bibliographystyle{IEEEtran}
\bibliography{IEEEabrv,IEEEexample}
	\vspace{-1cm}
\begin{IEEEbiographynophoto}
	{Yakun~Huang} is currently a Postdoctoral Researcher at the State Key Laboratory of Networking and Switching Technology, Beijing University of Posts and Telecommunications, Beijing, China.
	His current research interests include volumetric video streaming, mobile computing, and augmented reality.	
\end{IEEEbiographynophoto}
\vspace{-1cm}

\begin{IEEEbiographynophoto}
	{Yuanwei~Zhu} is currently working towards a Ph.D. degree at the State Key Laboratory of Networking and Switching Technology, Beijing University of Posts and Telecommunications, Beijing, China. His current research interests include point clouds, video streaming, and deep reinforcement learning.
\end{IEEEbiographynophoto}
\vspace{-1cm}

\begin{IEEEbiographynophoto}
	{Xiuquan~Qiao } is currently a Full Professor with the State Key Laboratory of Networking and Switching Technology, Beijing University of Posts and Telecommunications, Beijing, China. His current research interests include the future Internet, services computing, computer vision, distributed deep learning, augmented reality, virtual reality, and 5G networks.
\end{IEEEbiographynophoto}
\vspace{-1cm}

\begin{IEEEbiographynophoto}
	{Xiang~Su } is currently an Associate Professor with the Department of Computer Science, Norwegian University of Science and Technology, Norway, and the University of Oulu, Finland. He has extensive expertise in the Internet of Things, edge computing, mobile augmented reality, knowledge representations, and context modeling and reasoning. 
\end{IEEEbiographynophoto}
\vspace{-1cm}

\begin{IEEEbiographynophoto}
	{Schahram~Dustdar} (Fellow, IEEE) is a Full Professor of Computer Science and is heading the Distributed Systems Research Division at the TU Wien. He is an ACM Distinguished Scientist, ACM Distinguished Speaker, IEEE Fellow, and Member of Academia Europaea.
\end{IEEEbiographynophoto}
\vspace{-1cm}

\begin{IEEEbiographynophoto}
	{Ping~Zhang} (Fellow, IEEE) is currently a Full Professor and Director of the State Key Laboratory of Networking and Switching Technology, Beijing University of Posts and Telecommunications, Beijing, China.
	He is an Academician with the Chinese Academy of Engineering (CAE).
	He is also a member of the IMT-2020 (5G) Experts Panel and the Experts Panel for China’s 6G development.
\end{IEEEbiographynophoto}
\end{document}